\documentclass[sigconf]{acmart}
\AtBeginDocument{%
  \providecommand\BibTeX{{%
    \normalfont B\kern-0.5em{\scshape i\kern-0.25em b}\kern-0.8em\TeX}}}

\copyrightyear{2024}
\acmYear{2024}
\setcopyright{acmlicensed}
\acmConference[WSDM '24] {Proceedings of the 17th ACM International Conference on Web Search and Data Mining}{March 4--8, 2024}{Merida, Mexico.}
\acmBooktitle{Proceedings of the 17th ACM International Conference on Web Search and Data Mining (WSDM '24), March 4--8, 2024, Merida, Mexico}
\acmPrice{15.00}
\acmISBN{979-8-4007-0371-3/24/03}
\acmDOI{10.1145/3616855.3635780}

\usepackage{enumitem}
\usepackage{verbatim}
\usepackage{booktabs}
\usepackage{colortbl}
\usepackage{algorithm}
\usepackage{algorithmicx}
\usepackage{algpseudocode}
\usepackage{bbm}
\usepackage{array}
\usepackage{amsmath}
\usepackage{graphicx}
\usepackage{newfloat}
\usepackage{listings}
\usepackage{hyperref}
\usepackage{graphicx}
\usepackage{subfig}
\newcommand{\Set}[1]{\mathcal{#1}}
\newcommand{\ie}{\textit{i.e., }}
\newcommand{\eg}{\textit{e.g., }}

\newcommand{\cf}{\textit{cf. }}
\newcommand{\aka}{\textit{aka., }}
\newcommand{\cred}[1]{{\color{myred}{#1}}}
\newcommand{\cgreen}[1]{{\color{mygreen2}{#1}}}
\usepackage{multirow}
\usepackage{makecell}
\usepackage{diagbox}
\usepackage{subcaption}
\usepackage{stfloats}
\usepackage{pifont}
\definecolor{mygreen}{RGB}{0, 142, 0}
\definecolor{mygreen2}{RGB}{0, 180, 0}
\definecolor{myred}{RGB}{220, 0, 0}

\newtheorem{definition}{Definition}
\newtheorem{problem}{Problem}

\begin{document}
\title{Dynamic Sparse Learning: A Novel Paradigm for \\ Efficient Recommendation}

\author{Shuyao Wang}
\affiliation{%
  \institution{School of Data Science, University of Science and Technology of China}
  \country{}}
\email{shuyaowang@mail.ustc.edu.cn}

\author{Yongduo Sui}
\affiliation{%
  \institution{School of Data Science, University of Science and Technology of China}
  \country{}}
\email{syd2019@mail.ustc.edu.cn}

\author{Jiancan Wu}
\affiliation{%
  \institution{School of Information Science and Technology, University of Science and Technology of China}
  \country{}}
\email{wujcan@gmail.com}

\author{Zhi Zheng}
\affiliation{%
  \institution{School of Data Science, University of Science and Technology of China}
  \country{}}
\email{zhengzhi97@mail.ustc.edu.cn}

\author{Hui Xiong}
\authornote{Hui Xiong is the corresponding author.}
\affiliation{%
  \institution{The Thrust of Artificial Intelligence,
The Hong Kong University of Science
and Technology (Guangzhou)}
    \institution{The Department of Computer Science
and Engineering, The Hong Kong
University of Science and Technology}
\country{}}
\email{xionghui@ust.hk}

\renewcommand{\shortauthors}{Shuyao Wang, Yongduo Sui, Jiancan Wu, Zhi Zheng, \& Hui Xiong}
\begin{abstract}

In the realm of deep learning-based recommendation systems, the increasing computational demands, driven by the growing number of users and items, pose a significant challenge to practical deployment. This challenge is primarily twofold: reducing the model size while effectively learning user and item representations for efficient recommendations. Despite considerable advancements in model compression and architecture search, prevalent approaches face notable constraints. These include substantial additional computational costs from pre-training/re-training in model compression and an extensive search space in architecture design. Additionally, managing complexity and adhering to memory constraints is problematic, especially in scenarios with strict time or space limitations.
Addressing these issues, this paper introduces a novel learning paradigm, Dynamic Sparse Learning (DSL), tailored for recommendation models. DSL innovatively trains a lightweight sparse model from scratch, periodically evaluating and dynamically adjusting each weight's significance and the model's sparsity distribution during the training. This approach ensures a consistent and minimal parameter budget throughout the full learning lifecycle, paving the way for ``end-to-end'' efficiency from training to inference. Our extensive experimental results underline DSL's effectiveness, significantly reducing training and inference costs while delivering comparable recommendation performance.
We give an code link of our work: \url{https://github.com/shuyao-wang/DSL}.

\end{abstract}

\keywords{Efficient Recommendation, Sparse Learning}

\begin{CCSXML}
<ccs2012>
   <concept>
       <concept_id>10002951.10003227.10003351.10003269</concept_id>
       <concept_desc>Information systems~Collaborative filtering</concept_desc>
       <concept_significance>300</concept_significance>
       </concept>
   <concept>
       <concept_id>10002951.10003317.10003359.10003363</concept_id>
       <concept_desc>Information systems~Retrieval efficiency</concept_desc>
       <concept_significance>500</concept_significance>
       </concept>
   <concept>
       <concept_id>10002951.10003317.10003338.10003343</concept_id>
       <concept_desc>Information systems~Learning to rank</concept_desc>
       <concept_significance>500</concept_significance>
       </concept>
 </ccs2012>
\end{CCSXML}

\ccsdesc[500]{Information systems~Learning to rank}
\ccsdesc[500]{Information systems~Retrieval efficiency}
\ccsdesc[300]{Information systems~Collaborative filtering}

\maketitle

\section{Introduction}
Recommendation systems~\cite{zheng2022cbr,wang2020setrank,wang2021personalized} provide personalized services for today's web and have achieved significant success in various fields, such as e-commerce platforms, medical care~\cite{zheng2022ddr,zheng2023interaction,zheng2021drug}, education~\cite{zheng2023generative} and job search~\cite{zhi2023generative}.
At its core is learning high-quality representations for users and items based on historical interaction data. 
The rapidly growing population of users and items on one hand promotes the demands for large-scale recommender systems~\cite{wu2023survey}, while on the other hand posing great challenges for model training and inference, such as unaffordable memory overheads, computational costs, and inference latency.
Hence, efficient recommendation, \ie lightweight model, with low costs of training and inference, is of great need and significance.

\begin{table*}[t]
\centering
\caption{Comprehensive comparisons with existing efficient recommendation methods.}
\label{table:Method}
\vspace{-4mm}
\resizebox{0.8\textwidth}{!}{\begin{tabular}{l|cccccccc}
\toprule
\multirow{2}{*}{Method} & \multirow{2}{*}{Category}  & \multicolumn{3}{c}{Saving} & \multirow{2}{*}{End-to-end} & \multirow{2}{*}{\makecell[c]{Budget \\ controllable}} & \multirow{2}{*}{\makecell[c]{Non-uniform \\ embedding size}} & \multirow{2}{*}{\makecell[c]{Performance \\ reserved}} \\
\cline{3-5}
& & Training cost & Inference cost & Memory & & \\
\toprule
TKD  \cite{kang2021topology}      & KD        & \ding{55} & \ding{51} & \ding{55} & \ding{55} & \ding{51} & \ding{55} & \ding{55} \\
RKD \cite{tang2018ranking} 		  & KD        & \ding{55} & \ding{51} & \ding{55} & \ding{55} & \ding{51} & \ding{55} & \ding{55} \\
AutoEmb \cite{zhaok2021autoemb}   & AutoML    & \ding{55} & \ding{51} & \ding{55} & \ding{51} & \ding{55} & \ding{51} & \ding{51} \\
ESAPN \cite{liu2020automated}     & AutoML    & \ding{55} & \ding{51} & \ding{55} & \ding{51} & \ding{55} & \ding{51} & \ding{51} \\
PEP \cite{liu2021learnable}       & MP        & \ding{51} & \ding{51} & \ding{51} & \ding{51} & \ding{55} & \ding{51} & \ding{55} \\ 
LTH-MRS \cite{wang2022exploring}  & MP        & \ding{55} & \ding{51} & \ding{55} & \ding{55} & \ding{55} & \ding{51} & \ding{51} \\ 
\midrule
Ours     & MP & \ding{51} & \ding{51}  & \ding{51} & \ding{51} & \ding{51} & \ding{51} & \ding{51} \\
\bottomrule
\end{tabular}}
\vspace{-2mm}
\end{table*}

Various solutions have been proposed for efficient recommendation, which can be categorized into the following three research lines, each of which has inherent limitations.
\begin{itemize}[leftmargin=*,topsep=2.5mm]
\item \textbf{Knowledge Distillation (KD)} \cite{gou2021knowledge,kweon2021bidirectional,yang2022cross,tao2022revisiting,lee2021dual,kang2021topology,kang2022personalized} distills knowledge from a pre-trained large-scale model (\ie teacher model) to a compact model (\ie student model) for efficient inference. 
On the one hand, the proceeds of KD mainly lie in the reduction of inference costs by using the lightweight student model. 
Yet, it still needs to train a cumbersome teacher model from scratch, leaving the overall training cost not reduced.
On the other hand, the knowledge of the teacher models often cannot thoroughly transfer to the student models \cite{tang2018ranking}, usually incurring a large performance degradation \cite{kang2021topology,kweon2021bidirectional}. 

\item \textbf{Automated Machine Learning (AutoML)} \cite{automlsurvey,zheng2022automl,liu2021learnable,zhao2021autodim,zhaok2021autoemb,liu2020automated} aims to search lightweight model architectures in a given search space via diverse strategies, including gradient-based optimization \cite{ruder2016overview,liu2018darts}, reinforcement learning \cite{liu2021learnable} , or evolutionary algorithms \cite{qin2008differential}.
However, the search space should be manually predefined with human prior knowledge \cite{automlsurvey}. 
In addition, the complex procedures of optimization and performance estimation \cite{automlsurvey} also lead to an unaffordable computational cost.

\item \textbf{Model Pruning (MP)} \cite{frankle2018lottery,han2015deep,zhu2017prune,tanaka2020pruning,wang2022exploring} trims down the model size by removing redundant parameters from the full model.
It aims to find a sparse and lightweight model that can best retain the performance of the full model.
However, it needs to empirically find intriguing sparse embedding tables by an iterative ``train-prune-retrain'' pipeline \cite{frankle2018lottery,wang2022exploring}, which still suffers from the expensiveness of the post-training pruning.
\end{itemize}
% We select several representative efforts and summarize their properties in Table \ref{table:Method}.
% And we put more discussions about them in Section \ref{sec:rw}.
We present a clear summary of several representative efforts and their respective properties in Table~\ref{table:Method}. 
Further discussions can be found in Section~\ref{sec:rw}.
Scrutinizing the limitations of these solutions, they either require model pre-training or complex optimization processes of architecture search.
Hence, the benefits mainly come from the inference stage, while greatly increasing the training workload.

In view of that, we aim to design a simple and lightweight learning paradigm for recommendation models that can simultaneously trim down the training and inference costs, with comparable performance.
By inspecting the design of the conventional models, we find that most of them impose a constraint on the length of representations during model training, that is, embedding each user or item as a dense vector with a preset dimension (\aka embedding table).
However, in expectation, a flexible model is capable of assigning different embedding sizes to diverse users or items based on the information they carry.
Intuitively, users having multiple interests (or items attracting diverse audiences) should be more informative than the inactive counterparts, and hence should be represented with larger dimensions, and vice versa \cite{liu2020automated,zhaok2021autoemb, chen2023unbiased}.
Hence, there may exist a large number of redundant weights in the full embedding table, resulting in an over-parameterized model, which greatly hinders efficiency.
Then a question raises naturally: ``\textit{Can we relax the limitation on the embedding size of the model, and let the model automatically remove or add weights to achieve a dynamic embedding size during training?}''

Towards this end, we propose an end-to-end learning paradigm for recommendation: \underline{D}ynamic \underline{S}parse \underline{L}earning (\textbf{DSL}).
Specifically, given a randomly initialized model, DSL first randomly prunes the model to a predefined budget and trains the sparse model.
During the training process, it periodically adjusts the sparsity distribution of the model parameters via two dynamic strategies: pruning and growth.
The implemented pruning strategy identifies and eliminates weights that have negligible impact on the performance, thereby effectively reducing redundancy in the model parameters.
% The pruning strategy mines which weights will degrade the performance, thereby removing the redundant parameters.
While the growth process explores the potential informative weights that can improve the performance, thereby reactivating important parameters.
DSL sticks to a fixed and small budget by maintaining the same ratio of pruning and growth throughout the whole training stage, thus effectively reducing both the training and inference complexity.
DSL is a model-agnostic and plug-and-play learning framework that can be applied to various models.
In contrast to existing efforts in Table \ref{table:Method}, it achieves the attractive prospect of “end-to-end” efficiency from training to inference.
We implement DSL on diverse collaborative recommendation models, and conduct extensive experiments on benchmark datasets.
Experimental results show that DSL can effectively trim down both the training and inference costs, with comparable performance.

Overall, we make the following contributions:
\begin{itemize}[leftmargin=*,topsep=2.5mm]
% \setlength{\itemsep}{0pt}
% \setlength{\parsep}{0pt}
% \setlength{\parskip}{0pt}
% \vspace{-4mm}
\item \textbf{Problem}: We argue that large-scale recommendation models usually suffer from high training and inference complexity. 
Unfortunately, most solutions mainly alleviate the computational cost of inference, but double the cost of model training.

\item \textbf{Algorithm}: We propose an efficient learning paradigm for recommendation, named DSL. 
It trains lightweight sparse models and sticks to a fixed parameter budget during the whole training stage.
Hence, it can achieve end-to-end efficiency from training to inference.
%  thereby effectively trimming down both the training and inference costs.

\item \textbf{Experiments}:  We conduct extensive experiments on diverse recommendation models. 
The results demonstrate the superiority and effectiveness of DSL. More visualizations with in-depth analyses demonstrate the rationality of DSL.
\end{itemize}

\section{Preliminaries}

\subsection{Notations}
This paper focuses on collaborative recommendation settings.
Given a recommender system with $N$ users and $M$ items, we define the set of users $\mathcal{U}=\left\{u\right\}$, items $\mathcal{I}=\{i\}$, and their interactions $\mathcal{{O}} = {\left\{y_{ui}\right\}}$, where $y_{ui}=1$ denotes user $u$ has adopted item $i$ before, otherwise $y_{ui}=0$.
For convenience, we organize them as a graph $\mathcal{G}=\left\{\mathcal{V},\mathcal{E}\right\}$.
where $\mathcal{V}=\left\{v_{1}, \cdots, v_{N+M}\right\}$ is the node set comprising both user and item nodes, $\mathcal{E}=\left\{(u,i) | u\in\mathcal{U}, i\in\mathcal{I}, y_{ui}=1 \right\}$ is the edge set containing all observed interactions between users and items.
% And we use the node set $\mathcal{V}=\left\{v_{1}, \cdots, v_{N+M}\right\}$ to describe all users and items, where $\mathcal{V}=\mathcal{U} \cup \mathcal{I}$.
% We use the edge set $\mathcal{E}=\left\{(u,i) | u\in\mathcal{U}, i\in\mathcal{I}, y_{ui}=1 \right\}$ to denote the interactions between users and items.
Each node is encoded into a $d$-dimensional embedding vector $\mathbf{e} \in \mathbbm{R}^{d}$. 
We define $f$ as the recommendation model, and the embedding table of $f$ can be represented as follow:
\begin{equation} 
\mathbf{E} = [\underbrace{\mathbf{e}_{u_1}, \mathbf{e}_{u_2}, \cdots, \mathbf{e}_{u_N}}_{\text {users}}; \underbrace{\mathbf{e}_{i_1}, \mathbf{e}_{i_2}, \cdots, \mathbf{e}_{i_M}}_{\text{items}}]^{\top} \in \mathbb{R}^{(N+M) \times d}.
\end{equation}
% After model training, collaborative recommendation utilizes embedding vectors to achieve predictions for users or items.
% After training,  recommendation model will generate recommendation list based on learned user and item representations.
For a given recommendation model $f$, it's observable that the parameters of the embedding table expand linearly with the number of users and items. In collaborative recommendation, popular models like NeuMF \cite{he2017neural} and LightGCN \cite{he2020lightgcn} primarily rely on learning the embedding table.
Though other learnable parameters in $f$ may impact the final computational cost, this paper primarily considers the cost associated with embeddings.

\subsection{Efficient Recommendation}

% Efficiency generally refers to the ability to achieve model training or inference with fewer resources while maintaining performance comparable to the original model.
% In this work, we investigate the efficiency in the context of recommendation.

\subsubsection{\textbf{Problem Formulation.}}

The efficient recommendation aims to conserve resources while maintaining the recommendation quality. These resources often encompass computational costs during training and testing, memory space, and latency time, all crucial for practical applications. We now formally define the problem of efficient recommendation.
\begin{problem}[Efficient Recommendation]\label{pro:1}
Given a well-trained recommendation model $f$ that attains performance level $P$, which requires resources $C_{\rm tr}$, $C_{\rm te}$ and $C_{\rm m}$ for training, testing, and memory cost, respectively. Consider another model $f'$ that achieves performance level $P'$ after convergence with $C_{\rm tr}'$, $C_{\rm te}'$ and $C_{\rm m}'$. The task of efficient recommendation is to find the alternative $f'$ that is capable of replacing $f$, satisfying: $C_{\rm tr}' < C_{\rm tr}$, $C_{\rm te}' < C_{\rm te}$, $C_{\rm m}' < C_{\rm m}$ and $P'\approx P$.
\end{problem}
% Examining Problem \ref{pro:1}, it requires us to look for an alternative model with lightweight parameters or size to save resources. Coarse-grained and fine-grained
% Existing methods can compress the model from the model macrostructure, such as the dimension of the embedding table.
% While from a microscopic view, other efforts make fine-grained compression to sparsify the weight of the model.
The objective of Problem \ref{pro:1} is to look for a model with fewer parameters to conserve resources. 
We summarize and discuss existing methods towards this goal from two perspectives: coarse-grained model architectures and fine-grained model architectures.
\subsubsection{\textbf{Efficiency via Coarse-grained Model Architectures.}}

The efficiency of the model is influenced by its coarse-grained architecture, which is determined by the dimensions of the embedding table. 
Reducing the dimension of the embedding table can enhance efficiency for model training and inference.
% The model's coarse-grained architecture, dictated by the embedding table's dimensions, impacts efficiency.
% Training with lower dimensions can improve efficiency.
However, these streamlined models often struggle to deliver the promising performance equivalent to their larger counterparts, \ie $P'\approx P$ in Problem \ref{pro:1} does not hold.
To address this issue, knowledge distillation (KD) methods are employed \cite{tang2018ranking,kang2021topology,kweon2021bidirectional}, which distills knowledge from larger pre-trained teacher models into smaller ones to improve the recommendation performance. 
However, extensive studies \cite{kweon2021bidirectional} demonstrate that there still exists a non-negligible performance gap between small and large models.
Furthermore, the computational expense of pre-training is not mitigated, rendering these approaches insufficient for fully resolving Problem \ref{pro:1}.

\subsubsection{\textbf{Efficiency via Fine-grained Model Architectures.}}
Approaches from a microscopic perspective assign distinct embedding dimensions to individual users and items, presenting a fine-grained model design philosophy to achieve efficiency.
AutoML-based methods \cite{zheng2022automl,automlsurvey} search for the optimal model architecture within a predefined search space.
However, perfectly defining the search space is challenging \cite{zheng2022automl}.
Furthermore, complex procedures of optimization and performance estimation \cite{automlsurvey} also double the cost, making it difficult to solve Problem \ref{pro:1}.
Another research line is model pruning, guided by the lottery ticket hypothesis (LTH) \cite{frankle2018lottery,chen2020gans,JCST-2206-12583}. 
It states that sparse models discovered by pruning can replace full models without performance degradation.
Despite its potential benefits, LTH has not been extensively studied in the field of recommender systems. In this regard, we provide a formal definition of the winning ticket in recommendation.

\begin{definition}[Winning Ticket]\label{def:1}
Given a recommendation model $f$ with embedding table $\mathbf{E}$, a binary mask $\mathbf{M} \in \left\{0,1\right\}^{{\Vert \mathbf{E} \Vert}_0}$ can create a sparse embedding table: $\mathbf{E}' \leftarrow \mathbf{M} \odot \mathbf{E}$, where $\odot$ denotes the element-wise product.

1. The model $f(\mathcal{G}, \mathbf{E})$ with a dense embedding table $\mathbf{E}$ can obtain the performance $P$ after training $T$ iterations.

2. A sparse model $f(\mathcal{G},\mathbf{E}')$ can obtain the performance level $P'$ after training $T'$ iterations.

If $\exists \, \mathbf{M} \in \left\{0,1\right\}^{{\Vert \mathbf{E} \Vert}_0}$ such that $T' \leq T$, $P' \geq P$ and ${{\Vert \mathbf{E}' \Vert}_0} \ll {{\Vert \mathbf{E} \Vert}_0}$, then we define the sparse model $\mathbf{E}'$ as a winning ticket.

\end{definition}

% As per Definition \ref{def:1}, we can find that winning tickets can solve Problem \ref{pro:1} naturally.
As per Definition \ref{def:1}, winning tickets present a natural solution to Problem \ref{pro:1}, which possesses an attractive property: we could have trained from a sparse lightweight model if only we had known which mask to choose.
Recent work LTH-MRS \cite{wang2022exploring} has explored and validated the existence of lottery tickets in recommendation models.
However, it adopts a similar pruning strategy as in LTH \cite{frankle2018lottery} --- iterative magnitude pruning --- to locate the winning tickets.
This multiple-round training strategy is a surefire way to find the winning ticket, while it will largely increase the training costs. 
Although winning tickets are fascinating, how to efficiently find them is still an open problem \cite{you2019drawing,evci2020gradient}.
We summarize some specific studies of efficient recommendation in Table \ref{table:Method}.
More discussions are provided in Section \ref{sec:rw}.

\section{Methodology}

\begin{figure}[t]
\centering
\includegraphics[width=1\linewidth]{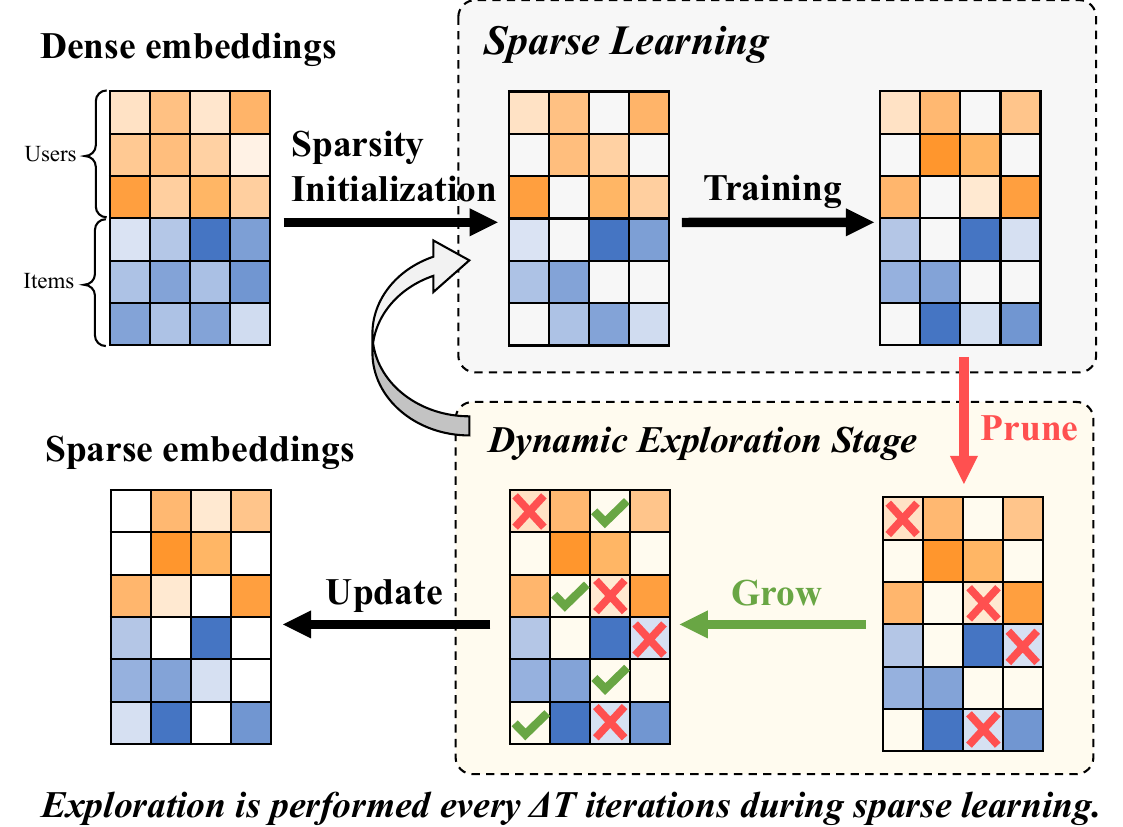}
\caption{The overview of the proposed Dynamic Sparse Learning (DSL) framework.}
\label{fig:model}
\vspace{-4mm}
\end{figure}

\subsection{Dynamic Sparse Learning}

In conventional model learning, all users and items are represented by embeddings of uniform size.
However, since the amount of information carried by diverse users and items is very likely to be different, the strategy of assigning the same embedding size may be suboptimal.
Observations from prior studies such as ESAPN \cite{liu2020automated}, UnKD \cite{chen2023unbiased}, and LTH-MRS \cite{wang2022exploring}, suggest that inactive users and unpopular items may contain less information, necessitating smaller embedding sizes to depict their simpler characteristics.
% Smaller embedding sizes are enough to describe their simple attributes or interests.
Hence, we relax the constraint on embedding size and encourage the model to dynamically discover those redundant weights during learning.
% We name our proposed learning paradigm as \underline{D}ynamic \underline{S}parse \underline{L}earning (\textbf{DSL}).
The overview of DSL is displayed in Figure~\ref{fig:model}, which includes three steps: sparsity initialization, sparse learning, and dynamic exploration.
% DSL transitions between the sparse learning and the dynamic exploration stage multiple times.
DSL initiates the training process with a randomly sparsified model. Furthermore, the dynamic exploration step is periodically (\ie $\Delta{T}$ training iterations) introduced during the sparse learning process. 
Now we elaborate on these three steps in detail.

\subsubsection{\textbf{Sparsity Initialization}}
Given a dense user-item embedding table $\mathbf{E}$, we first establish the model's sparsity $s\in(0,1)$. Then we initialize a binary mask $\mathbf{M} \in \left\{0, 1\right\}^{{\Vert \mathbf{E} \Vert}_0}$ with a random sparsity distribution, where ${{\Vert \mathbf{M} \Vert}_0}={{\Vert \mathbf{E} \Vert}_0}\cdot(1-s)$, and we can obtain a sparse embedding table $\mathbf{E}' \leftarrow \mathbf{M} \odot \mathbf{E}$. 
Please note that the binary mask is solely introduced to aid in the description of our operations. In practice, there is no need to store this additional matrix.
To ensure a constant model parameter budget, we fix the sparsity $s$ throughout the whole training process.

\subsubsection{\textbf{Sparse Learning}}
We directly train the sparse model until entering the next dynamic exploration stage.
As shown in Figure \ref{fig:model}, we will transition between the exploration stage and the sparse learning multiple times.
Since a new sparse distribution, \ie model architecture, is obtained after each dynamic exploration, the model needs to retune parameters to fit new architecture.
We define $T_{end}$ as the total training iterations and $\Delta{T}$ as the number of iterations for each round of the sparse learning.
We find that both too-large and too-small values of $\Delta{T}$ can greatly deteriorate the performance.
Given a fixed training iteration ${T_{end}}$, if we make $\Delta{T}$ too large, it will greatly reduce the number of times entering the exploration step, which means that the model has fewer opportunities to modify the current model structure.
Ultimately, when $\Delta{T}\to\infty$, it is equivalent to directly training a randomly pruned model, which will lead to poor performance.
When $\Delta{T}$ is too small, the model cannot adequately adjust the parameters to adapt to the current architecture, so when entering the exploration stage, it is difficult to accurately remove redundant weights or add important weights.
This will undoubtedly degrade the performance.
Hence, an appropriate number of sparse training iterations $\Delta{T}$ is necessary.
In view of that, we conduct comprehensive ablation studies in Section \ref{sec:interval}.

\begin{algorithm}[t]
\renewcommand{\algorithmicrequire}{\textbf{Input:}}
\renewcommand{\algorithmicensure}{\textbf{Output:}}
\caption{Dynamic Sparse Learning}
\label{alg1}
\begin{algorithmic}[1]
\Require Dense embedding table $\mathbf{E}$, Dataset $\mathcal{D}$, 
Random binary mask $\mathbf{M} \in \left\{0, 1\right\}^{{\Vert \mathbf{E} \Vert}_0}$ with sparsity $s$, Loss $\mathcal{L}$, Learning rate $\eta$, Initial update ratio $\rho_0$, $\Delta{T}$, $T_{end}$.
\State Initialize sparse table $\mathbf{E} \gets \mathbf{M}\odot\mathbf{E}$
% \Comment{Highly reduced parameter count.}
% \For{each training iteration $t$}
\For{training iteration $t \in \{1, ..., T_{end}\}$}
\State Sampling a batch $b_t \sim \mathcal{D}$
\If{$(t\ {\rm mod} \ {\Delta{T}}==0)$}
% \State $\rho \gets f_{\rm{decay}}(t, \alpha, \rm{T_{end}})$
\State $\rho_t \gets {\rm CosineAnnealing}(\rho_0, t, T_{end})$
\State Creating $\mathbf{M}$ by pruning and growth with $\rho_t$  
% \State $\mathbf{E} \gets \mathbf{M}\odot\mathbf{E}$
\State Updating sparse embedding table $\mathbf{E}$
\Else \Statex\qquad\qquad{$\mathbf{E} \gets \mathbf{E} - {\eta} {\nabla_\mathbf{E}}{\mathcal{L}}$}
\EndIf
\EndFor\\
\Return A sparse embedding table

\end{algorithmic}  
\end{algorithm}

\subsubsection{\textbf{Dynamic Exploration}} 
Exploration is performed every $\Delta{T}$ iteration during the model training process. 
In this step, we dynamically adjust the sparsity distribution of the model parameters.
There exist three key components: update schedule, pruning principle, and growth principle.
\begin{itemize}[leftmargin=4mm]
\item \textbf{Update schedule.} 
To ensure the stability of the exploration stage, we set the update ratio $\rho_t$ for every exploration using a cosine annealing schedule, where $t$ is the current training step. Take $\rho_0$ as the initial update ratio at the 0-th iteration.
Following \cite{loshchilov2016sgdr}, the decay function is defined as:
\begin{equation} 
    {\rho_t} = \frac{\rho_0}{2}\left(1+ {\rm cos}\left(\frac{\pi t}{T_{\text {end }}}\right)\right).
\end{equation}
% \begin{equation} 
%     {\rho_t} =\rm{CosineAnnealing}({\rho_0}, t, \rm{T_{end}}) = \frac{\rho_0}{2}\left(1+cos\left(\frac{\pi t}{T_{\text {end }}}\right)\right).
% \end{equation}
As model training approaches convergence, the update ratio should gradually decrease to ensure stable convergence.
Hence, this decay strategy facilitates the gradual adoption of the model architecture to stabilize the training process.
We also verify the effectiveness of this strategy in Section~\ref{sec:update}.

\item \textbf{Pruning principle.}
Following \cite{frankle2018lottery}, we adopt the weight magnitude as the pruning indicator.
We prune a $\rho_t$ ratio of the lowest-magnitude weights. It is worth mentioning that we maintain the same update ratio in both the pruning and growth process to stick to a fixed budget.

\item \textbf{Growth principle.}
We monitor the gradients of the pruned weights to assess their potential importance in the final prediction. Then we reactivate a $\rho_t$ ratio of the pruned weights having the highest gradient magnitudes, which will participate in the next round of sparse learning. Such a growth principle provides a \textbf{regret mechanism} for pruning, serving as compensation for the information loss caused by the previous greedy pruning \cite{wang2022exploring}.

\end{itemize}
The overview of the proposed framework is depicted in Figure \ref{fig:model}, and the detailed pipeline of DSL is summarized in Algorithm \ref{alg1}.

\subsection{Technique Analysis}
In this section, we provide detailed analyses and discussions of the working mechanism of the DSL.
It can achieve sparse learning while retaining the performance for the following reasons:
\begin{itemize}[leftmargin=4mm]
\item \textbf{Dynamic Architecture Exploration.}
Recklessly training a static model architecture with random sparsity distribution often leads to suboptimal performance \cite{wang2022exploring}.
% Static sparse training adopts fixed model structures for training, which is highly likely to be suboptimal.
% This is because of existing poor natural model architectures \cite{liu2018darts} that lead to inferior performance.
This is mainly due to the existence of poor natural model architectures \cite{liu2018darts}, which result in subpar performance.
In contrast, DSL flexibly adjusts the suboptimal structure during training, allowing the model to dynamically learn the importance of weights, thus achieving better performance.
\item \textbf{Sufficient Sparse Learning.}
DSL monitors the states of the parameters, such as magnitudes or gradients, which reflect their importance.
This provides a guarantee for the subsequent precise parameter pruning and growth process.
However, it will also lead to insufficient exploration, resulting in a trade-off between the number of exploration cycles and sparse learning iterations.
In Section \ref{sec:ablation}, we conduct extensive experiments to verify this.
\item \textbf{Sufficient Exploration Space.}
One key strength of DSL is its ability to reactivate important parameters that were pruned, thus avoiding the problem of degraded performance in compressed models. Furthermore, as DSL searches for the most critical pruned parameters to reactivate in each round, DSL has the potential to surpass the original model's performance (\cf Section \ref{sec:baseline}). Additionally, the regret mechanism imbues DSL with more flexibility in adapting to different model architectures, thus enhancing the stability of the compressed model.
% The effective exploration parameters refer to newly grown parameters.
% It means that DSL can reactivate some important but pruned weights.
% If these parameters are sufficient, such as covering the entire parameter exploration space of the model, it is possible to achieve the best performance under the current sparsity.
% In our experiments, we found that the effective exploration parameters are almost equal to the full exploration space of the model.
\end{itemize}

\vspace{-2mm}
\subsection{Complexity Analysis}
% In this section, we provide a brief analysis of the time complexity of DSL.
% Without loss of generality, we take a popular graph-based model, LightGCN \cite{he2020lightgcn}, as an example.
We use LightGCN \cite{he2020lightgcn}, a widely used collaborative recommendation model, to instantiate this.
It contains two critical components: light graph convolution (LGC) and layer combination (LC).
Assuming that it performs $L$-layer LGC with the adjacency matrix $\mathbf{A}$ of graph $\Set{G}$.
Then the time complexity of LGC is $\mathcal{O}\left(\|\mathbf{A}\|_{0} \times d \times L \right)$; LC is $\mathcal{O}\left(N\times d + M \times d\right)$ and similarity computation is $\mathcal{O}\left(N\times M \times d \right)$. 
Due to $(N+M) \ll (N \times M)$, the total time complexity is approximated to $\mathcal{O}\left((\|\mathbf{A}\|_{0} \times L+N \times M) \times d \right)$. 
For the sparse model with sparsity $s$, the time complexity will be greatly reduced to $\mathcal{O}\left((\|\mathbf{A}\|_{0} \times L+N \times M) \times d \times (1-s)\right)$.

\section{Experiments}

\begin{table*}[t]
\centering
\caption{Performance over diverse models. 
$l$ and $d$ denote the layer number and the embedding size, respectively.
Following work \cite{chen2021unified}, we use MACs (Tr.) and  MACs (In.) to measure the training and inference computational costs, respectively.}
\label{table:backbone}
% \begin{threeparttable}
\vspace{-4mm}
\resizebox{\textwidth}{40mm}{\begin{tabular}{ccccccccccccc}
\toprule
\multirow{2}{*}{Model} & \multicolumn{4}{c}{MovieLens-1M} & \multicolumn{4}{c}{CiteUlike} & \multicolumn{4}{c}{Foursquare} \\ 
% \cmidrule(r){2-13}
\cmidrule(r){2-5}  \cmidrule(r){6-9} \cmidrule(r){10-13}
& MACs (Tr.) & MACs (In.) & Recall/HR & NDCG & MACs (Tr.) & MACs (In.) & Recall/HR & NDCG & MACs (Tr.) & MACs (In.) & Recall/HR & NDCG \\
\toprule
NeuMF                      & 3.13e+13 & 9.68e+08 & 0.7867 & 0.4478  & 3.85e+12 & 8.87e+08 & 0.4000 & 0.2053 & 3.72e+13 & 2.50e+09 & 0.4940 & 0.2723 \\
\rowcolor{gray!20} + DSL   & 2.09e+13 & 6.47e+08 & 0.7525 & 0.4531  & 2.58e+12 & 5.93e+08 & 0.4000 & 0.2809 & 2.49e+13 & 1.67e+09 & 0.5663 & 0.3449 \\
\midrule
ConvNCF                    & 1.50e+15 & 4.65e+10 & 0.7745 & 0.5261  & 1.83e+14 & 4.26e+10 & 0.3429 & 0.2816 & 1.76e+15 & 1.20e+11 & 0.5301 & 0.2583 \\
\rowcolor{gray!20} + DSL   & 0.38e+15 & 1.19e+10 & 0.7990 & 0.4875  & 0.46e+14 & 1.09e+10 & 0.3714 & 0.2076 & 0.45e+15 & 0.31e+11 & 0.4940 & 0.3115      \\
\midrule
MultVAE                    & 2.40e+12 & 1.08e+10 & 0.3099 & 0.3052  & 1.36e+13 & 6.40e+10 & 0.2260 & 0.1278 & 4.61e+13 & 2.24e+11 & 0.1550 & 0.1639 \\
\rowcolor{gray!20} + DSL   & 1.20e+12 & 0.54e+10 & 0.3080 & 0.3005  & 0.68e+13 & 3.20e+10 & 0.2044 & 0.1093 & 3.69e+13 & 1.79e+11 & 0.1427 & 0.1485 \\
\midrule
CDAE                       & 2.67e+11 & 1.26e+10 & 0.3254 & 0.3236  & 3.60e+12 & 6.77e+10 & 0.1687 & 0.1080 & 1.15e+13 & 2.27e+11 & 0.2504 & 0.2452 \\
\rowcolor{gray!20} + DSL   & 1.33e+11 & 0.63e+10 & 0.3657 & 0.3656  & 1.80e+12 & 3.39e+10 & 0.1423 & 0.0962 & 0.92e+13 & 1.82e+11 & 0.2350 & 0.2303 \\
\midrule
\multirow{2}{*}{} & \multicolumn{4}{c}{Amazon-Book} & \multicolumn{4}{c}{Yelp2018} & \multicolumn{4}{c}{Gowalla} \\ 
\cmidrule(r){2-5}  \cmidrule(r){6-9} \cmidrule(r){10-13}
& MACs (Tr.) & MACs (In.) & Recall & NDCG  & MACs (Tr.) & MACs (In.) & Recall & NDCG & MACs (Tr.) & MACs (In.) & Recall & NDCG \\
\midrule
LightGCN ($l$=3)           & 1.59e+20 & 6.88e+13 & 0.0414 & 0.0321 & 4.00e+19 & 3.32e+13 & 0.0642 & 0.0528 & 1.75e+19 & 2.21e+13 & 0.1816 & 0.1550  \\ 
\rowcolor{gray!20} + DSL   & 0.95e+20 & 4.14e+13 & 0.0416 & 0.0325 & 2.67e+19 & 2.22e+13 & 0.0641 & 0.0527 & 1.16e+19 & 1.48e+13 & 0.1813 & 0.1547  \\ 
\midrule
LightGCN ($l$=4)           & 2.12e+20 & 9.17e+13 & 0.0406 & 0.0313 & 5.34e+19 & 4.42e+13 & 0.0649 & 0.0530 & 2.33e+19 & 2.95e+13 & 0.1830 & 0.1550  \\ 
\rowcolor{gray!20} + DSL   & 1.17e+20 & 5.06e+13 & 0.0405 & 0.0314 & 3.34e+19 & 2.77e+13 & 0.0651 & 0.0536 & 1.45e+19 & 1.84e+13 & 0.1821 & 0.1544  \\ 
\midrule
UltraGCN ($d$=64)          & 5.02e+13 & 6.24e+11 & 0.0678 & 0.0553 & 2.37e+13 & 15.5e+10 & 0.0673 & 0.0554 & 9.05e+13 & 1.61e+11 & 0.1858 & 0.1576  \\ 
\rowcolor{gray!20} + DSL   & 2.69e+13 & 1.56e+11 & 0.0685 & 0.0563 & 1.68e+13 & 9.89e+10 & 0.0645 & 0.0530 & 6.23e+13 & 1.03e+11 & 0.1742 & 0.1448  \\ 
\midrule
UltraGCN ($d$=128)         & 9.85e+13 & 1.25e+12 & 0.0712 & 0.0582 & 4.62e+13 & 3.09e+11 & 0.0676 & 0.0556 & 1.76e+14 & 3.22e+11 & 0.1844 & 0.1539  \\ 
\rowcolor{gray!20} + DSL   & 5.18e+13 & 0.31e+12 & 0.0727 & 0.0582 & 3.23e+13 & 1.98e+11 & 0.0651 & 0.0532 & 1.20e+14 & 2.06e+11 & 0.1792 & 0.1478  \\
\bottomrule
\end{tabular}}

% \begin{tablenotes}
%         \footnotesize
%         \item[*]$l$ and $d$ denote the layer number and the embedding dimension, respectively. 
%         % \item[**] my website is ... 
%       \end{tablenotes}
%       \end{threeparttable}
\end{table*}

To verify the superiority of DSL, we conduct extensive experiments to answer the following \textbf{R}esearch \textbf{Q}uestions:

\begin{itemize}[leftmargin=4mm]
\item \textbf{RQ1:} How does DSL perform in terms of cost and performance, when applying to diverse recommendation models?
\item \textbf{RQ2:} Compared with the state-of-the-art solutions for efficient recommendation, how does DSL perform?
\item \textbf{RQ3:} For different hyper-parameters in DSL, what are their roles or impacts on performance?
\item \textbf{RQ4:} Does DSL explore the model architectures with regular patterns or insightful interpretations?
\end{itemize}

\subsection{Experimental Setup}

\subsubsection{\textbf{Datasets \& Metrics.}}
We conduct experiments on 6 benchmark datasets, including 3 small-scale datasets: MovieLens-1M, CiteUlike and Foursquare, and 3 large-scale datasets: Amazon-Book, Yelp2018 and Gowalla. 
According to problem definition \ref{pro:1}, we need to evaluate our method from the following two perspectives.
In terms of model performance, we adopt the full-ranking protocol \cite{he2020lightgcn} and present the results using the widely used Recall@20, HR@20, and NDCG@20 metrics. 
In terms of costs, following most pruning-related studies \cite{chensparsity}, we adopt the Training MACs (Tr.), Inference MACs (In.), and Memory to evaluate the computational efficiency and memory overhead. 
% In terms of costs, we adopt the Training MACs (Tr.), Inference MACs (In.), and Memory to evaluate the computational efficiency and memory overhead.
% We provide detailed data statistics in Appendix~\ref{apd:dataset} and metric descriptions in Appendix~\ref{apd:metric}.

\subsubsection{\textbf{Backbone Models.}}
% We adopt the popular models: LightGCN \cite{he2020lightgcn}, UltraGCN \cite{mao2021ultragcn} and NGCF \cite{wang2019neural} with diverse layers or embedding sizes.
To verify that DSL can be applied to diverse models, we choose 6 backbone models, which can be divided into the following three categories:
\begin{itemize}[leftmargin=4mm]
\item \textbf{Collaborative filtering models:} NeuMF \cite{he2017neural}, ConvNCF \cite{he2018outer}.
\item \textbf{Autoencoder-based models:} CDAE \cite{wu2016collaborative}, MultVAE \cite{liang2018variational}.
\item \textbf{Graph-based models:} LightGCN \cite{he2020lightgcn}, UltraGCN \cite{mao2021ultragcn}.
\end{itemize}

\subsubsection{\textbf{Baselines.}}
To demonstrate the superiority of DSL, we compare it with diverse state-of-the-art solutions for efficient recommendation. 
% All baseline methods are fine-tuned within the search range provided in the original paper.
They fall into the following three categories:
\begin{itemize}[leftmargin=4mm]
\item \textbf{KD-based methods:} TKD \cite{kang2021topology}, RKD \cite{tang2018ranking}.
\item \textbf{AutoML-based methods:} AutoEmb \cite{zhaok2021autoemb}, ESAPN \cite{liu2020automated}.
\item \textbf{MP-based methods:} PEP \cite{liu2021learnable}, LTH-MRS \cite{wang2022exploring}, Random Pruning (RP) \cite{zhu2017prune}, One-shot Magnitude Pruning (OMP) \cite{han2015deep}, Without Rewinding (WR) \cite{savarese2020winning}. 
% Random Pruning (RP) \cite{zhu2017prune,wang2022exploring} adopts one-shot random pruning before training and completes the entire training process with a fixed mask.
% One-shot Magnitude Pruning (OMP) \cite{han2015deep} performs one-shot pruning on lowest-magnitude weights after full model pre-training.
% Without Rewinding (WR) \cite{savarese2020winning} is a variant for iterative magnitude pruning~\cite{frankle2018lottery,wang2022exploring}. 
% It performs iterative magnitude pruning without rewinding the initialization of the original model.

\end{itemize}

\subsection{Performance Evaluations (RQ1)}\label{sec:backbone}
We implement DSL on various models, with settings detailed in the code link. 
% The results, found in Table \ref{table:backbone}, represent extreme sparsity levels, the highest level achievable without significant performance loss.
The results presented in Table \ref{table:backbone} highlight the achievement of extreme sparsity levels --- the highest degree of sparsity attainable without incurring significant performance loss.
From these results, we make the following \textbf{Obs}ervations:

\textbf{Obs1: DSL consistently reduces both training and inference costs on diverse models with comparable performance.}
DSL is first applied to four backbone models, \ie CF-based and autoencoder-based models, using three small-scale datasets (refer to the first four rows).
Specifically, with NeuMF and ConvNCF, the computational costs of training and testing are reduced by approximately 50.0\% $\sim$ 74.4\%. For autoencoder-based models, MultVAE and CDAE, the training and inference MACs are cut by about 20\% $\sim$ 50\%. These findings underline the effectiveness of DSL with classic models and smaller datasets.
Next, DSL is deployed on two widely used graph-based models of different layer depths and embedding sizes, and tested on three large-scale datasets.
% For LightGCN with different layers, DSL achieves 33.3\%$\sim$45.1\% training MACs saving and 33.2\%$\sim$44.9\% inference MACs saving.
% For UltraGCN with different embedding sizes, DSL achieves 29.3\%$\sim$47.4\% training MACs saving and 36.1\%$\sim$75.0\% inference MACs saving, with minor performance drops.
For LightGCN with different layers and UltraGCN with different embedding sizes, DSL achieves 29.3\%$\sim$47.4\% training MACs saving and 33.2\%$\sim$75.0\% inference MACs saving.
These findings confirm the efficacy of DSL in reducing both training and inference costs across diverse models and datasets without significant performance compromise. 
Moreover, DSL outperforms the full model in some cases, such as NeuMF on MovieLens-1M and CiteUlike, and UltraGCN on the Amazon-Book. This suggests DSL's ability to prune redundant weights and explore more meaningful model architectures.

\begin{table*}[t]
\centering
\caption{Result comparisons of diverse solutions for efficient recommendation. The average and standard deviation results are reported across five random runs. We adopt the models with different embedding sizes. The bold and underlined numbers highlight the best and the second best performance, respectively.}
\vspace{-4mm}
\label{table:baselines}
\resizebox{\textwidth}{28mm}{\begin{tabular}{ccccccccccc}
\toprule
\multirow{2}{*}{Category} & \multirow{2}{*}{Method} & \multicolumn{4}{c}{Embedding size=64} & \multicolumn{4}{c}{Embedding size=128} & \multirow{2}{*}{\makecell[c]{Average \\ MACs}} \\ 
\cmidrule(r){3-6}  \cmidrule(r){7-10} &
& MACs (Tr.) & Memory & Recall & NDCG & MACs (Tr.) & Memory & Recall & NDCG &  \\
\toprule
& Baseline & 4.00e+19 & 2103M & 0.0642\scriptsize{(.0001)} & 0.0528\scriptsize{(.0002)} & 8.01e+19 & 2399M & {0.0673\scriptsize\scriptsize{(.0003)}} & 0.0550\scriptsize{(.0001)} & 0 \\ 
\midrule
\multirow{2}{*}{KD} & RKD  & 6.00e+19 & 2103M & 0.0558\scriptsize{(.0005)} & 0.0460\scriptsize{(.0004)} & 1.20e+20 & 2399M & 0.0610\scriptsize{(.0003)} & 0.0493\scriptsize{(.0004)} & \cred{+ 49.9\%}  \\
& TKD       & 6.67e+19 & 2103M & 0.0615\scriptsize{(.0007)} & 0.0514\scriptsize{(.0006)} & 1.34e+20 & 2399M & 0.0645\scriptsize{(.0005)} & 0.0533\scriptsize{(.0003)} & \cred{+ 67.1\%}  \\
\midrule
\multirow{2}{*}{AutoML} & AutoEmb & 7.92e+19 & 4210M & 0.0627\scriptsize{(.0002)} & 0.0511\scriptsize{(.0003)}  & 1.58e+20 & 4521M & 0.0654\scriptsize{(.0003)} & 0.0536\scriptsize{(.0003)} & \cred{+ 97.7\%} \\
& ESAPN  & 1.91e+20 & 4364M & \underline{0.0638\scriptsize{(.0002)}} & \underline{0.0525\scriptsize{(.0002)}} & 3.81e+20  & 4675M & \underline{0.0672\scriptsize{(.0003)}} & \underline{0.0545\scriptsize{(.0002)}} & \cred{+ 376.6\%}  \\
\midrule
\multirow{6}{*}{MP} & RP  & \textbf{2.67e+19}  & \textbf{1052M}  & 0.0557\scriptsize{(.0011)} & 0.0458\scriptsize{(.0010)} & \textbf{5.34e+19} & \textbf{1199M} & 0.0608\scriptsize{(.0009)} & 0.0498\scriptsize{(.0009)} & \textbf{\cgreen{- 33.3\%}} \\ 
& OMP      & 6.67e+19 & 2103M & 0.0622\scriptsize{(.0000)} & 0.0509\scriptsize{(.0001)} & 1.33e+20  & 2399M & 0.0661\scriptsize{(.0001)} & 0.0542\scriptsize{(.0000)} & \cred{+ 66.4\%}  \\ 
& WR       & 1.93e+20 & 2103M & 0.0621\scriptsize{(.0003)} & 0.0501\scriptsize{(.0004)} & 3.85e+20  & 2399M & 0.0661\scriptsize{(.0004)} & 0.0537\scriptsize{(.0005)} & \cred{+ 381.6\%}  \\ 
& PEP      & \underline{3.34e+19} & \underline{2082M} & 0.0592\scriptsize{(.0013)} & 0.0484\scriptsize{(.0011)} & \underline{6.68e+19} & \underline{2375M} & 0.0623\scriptsize{(.0009)} & 0.0501\scriptsize{(.0010)} & \cgreen{- 16.5\%}  \\
& LTH-MRS  & 1.93e+20 & 2103M  & 0.0635\scriptsize{(.0000)} & 0.0525\scriptsize{(.0001)} & 3.85e+20  & 2399M & 0.0666\scriptsize{(.0001)} & 0.0545\scriptsize{(.0001)} & \cred{+ 381.6\%}  \\ 
& DSL (Ours)   & \textbf{2.67e+19} & \textbf{1052M} & \textbf{0.0641\scriptsize{(.0002)}} & \textbf{0.0527\scriptsize{(.0001)}} & \textbf{5.34e+19} & \textbf{1199M} & \textbf{0.0675\scriptsize{(.0003)}} & \textbf{0.0553\scriptsize{(.0002)}} & \textbf{\cgreen{- 33.3\%}} \\ 
\bottomrule
\end{tabular}}
\end{table*}

\begin{figure*}[htbp]
\centering
\includegraphics[width=0.70\linewidth]{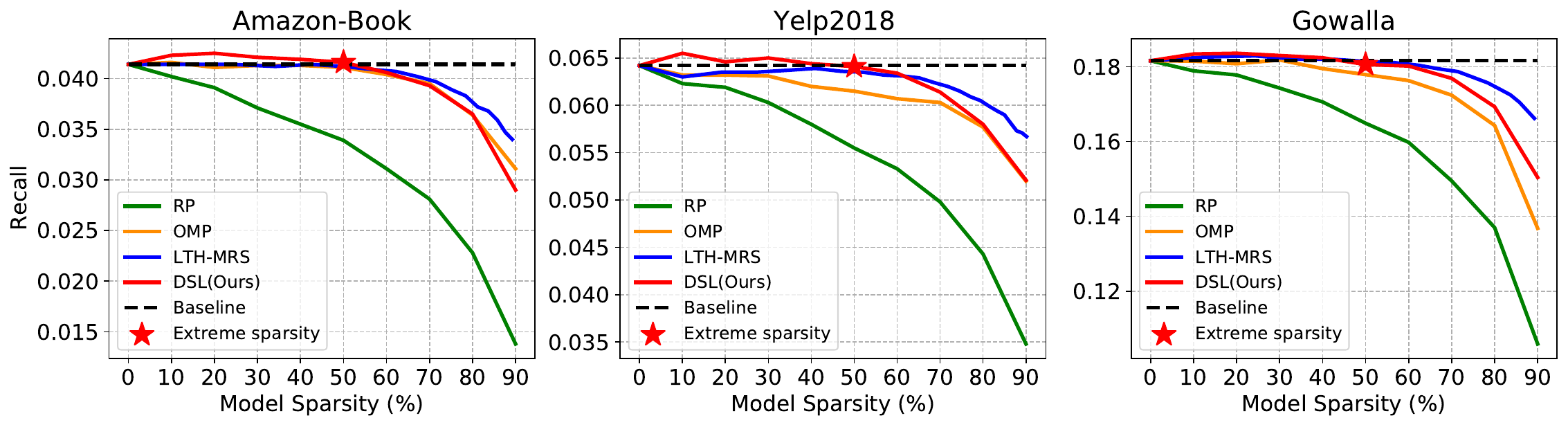}
\vspace{-4mm}
\caption{Performance comparisons with different sparsity levels. The star denotes the extreme sparsity of DSL, which achieves the similar performance levels as the baseline.}
\label{fig:sparse_recall}
\vspace{-4mm}
\end{figure*}

% The experimental results demonstrate that the DSL is suitable for diverse models with different layers and embedding sizes. 
% Specifically, for LightGCN with different layers, DSL achieves 33.3\%$\sim$45.1\% training MACs saving and 33.2\%$\sim$44.9\% inference MACs saving.
% For UltraGCN with different embedding sizes, DSL achieves 29.3\%$\sim$47.4\% training MACs saving and 36.1\%$\sim$75.0\% inference MACs saving, with minor performance drops.
% These results verify that DSL can effectively reduce both training and inference costs for diverse models with comparable performance.
% Specially, DSL even outperforms the full model in some cases, such as UltraGCN on Amazon-Book dataset and LightGCN ($l$=4) on Yelp2018 dataset.
% It denotes that DSL prunes redundant weights and explores more significant structures for backbone models.

\vspace{-2mm}
\subsection{Comparisons with Baselines (RQ2)}\label{sec:baseline}
We compare DSL with other baselines using LightGCN as backbone model, analyzing its performance across diverse embedding sizes. To ensure fairness, we commence by utilizing identical dense models and maintaining equivalent compression ratios. Specifically, we implement a 50\% sparsity level for pruning-based methods, whereas for KD-based methods, the size of the student models is half that of their corresponding teacher models, \ie the original dense models.
The experimental results are shown in Table \ref{table:baselines}, we make the following observations:
% The results, found in Table \ref{table:baselines}, yielded the following \textbf{Obs}ervations:

\textbf{Obs2: Both KD-based and AutoML-based methods suffer from large training costs.} 
For KD-based methods, the performance of RKD and TKD dropped on average by 11.22\% and 4.34\%, respectively.
Furthermore, they all require large-scale pre-trained teacher models, which greatly increases the training costs.
For RKD and TKD, the training MACs increase on average by 49.9\% and 67.1\%, respectively.
For AutoML-based methods, we can observe that both AutoEmb and ESAPN can keep comparable performance with baseline.
However, they also involve additional parameters, such as a controller or policy networks, and experience complex optimization processes. 
It inevitably results in large training computational and memory costs.
In contrast, our proposed DSL dynamically trains sparse and lightweight models.
On the one hand, its dynamic exploration stage can find suitable model architectures, thus guaranteeing comparable performance.
On the other hand, the sparse models and end-to-end training strategy guarantee low training computational costs and memory overhead.

\textbf{Obs3: DSL consistently outperforms other pruning-based methods.}
PEP achieves smaller training costs due to its dynamic training process. 
However, the performance drops by 6.84\%$\sim$9.03\% compared to the baseline, since the dynamic pruning thresholds will lead to unstable training.
OMP has relatively good performance, while it needs one-shot pruning on pre-trained models. 
In contrast, LTH-WRS outperforms OMP. 
The pruning ratio in each round is much smaller than OMP, so redundant weights can be pruned more accurately. 
Unfortunately, the iterative "train-prune-retain" pipeline incurs significant training costs amounting to approximately 380\% increase in comparison to the baseline. 
Moreover, while RP delivers comparable memory and training costs to DSL, its performance deteriorates by 9.66\%$\sim$13.24\%. 
This is attributed to RP's use of solely static sparse models during training, without incorporating DSL's exploratory stage.
Finally, we can easily observe that our proposed DSL can overcome all the shortcomings of the above pruning methods. 
It effectively saves training and memory costs with comparable performance.

\textbf{Obs4: DSL can achieve better ``performance-sparsity'' trade-offs.}
To explore the relationships between performance and the model sparsity, we plot ``performance-sparsity'' curve from 0$\sim$90\% sparsity levels.
The results are depicted in Figure \ref{fig:sparse_recall}.
We can observe that the performance of RP drops sharply with increasing sparsity.
OMP performs worse than DSL, while LTH-MRS achieves comparable performance with DSL.
We also observe that DSL even outperforms baseline in some cases, such as 10\%$\sim$40\% sparsity levels on Amazon-Book and 0$\sim$40\% sparsity levels on Yelp2018.
It demonstrates that the DSL can discover better model structures than the baselines through dynamic explorations.
These results further demonstrate that our proposed DSL can achieve better trade-offs while keeping lower computational costs.

\subsection{Ablation Study (RQ3)}\label{sec:ablation}

% \begin{table}[t]
% \centering
% \caption{Comparisons with different decay functions.}
% \vspace{-2mm}
% \renewcommand\arraystretch{1.2}
% \label{table:decay}
% \resizebox{0.46\textwidth}{12mm}{\begin{tabular}{lcccccc}
% \toprule
% \multirow{2}{*}{Type}   & \multicolumn{2}{c}{Amazon-book} & \multicolumn{2}{c}{Yelp2018} & \multicolumn{2}{c}{Gowalla} \\
% \cmidrule(r){2-3} \cmidrule(r){4-5} \cmidrule(r){6-7}
%     & Recall & NDCG & Recall & NDCG & Recall & NDCG \\
% \toprule
% Cosine   & 0.0406 & 0.0316 & 0.0641 & 0.0527 & 0.1812 & 0.1546  \\ 
% Linear   & 0.0409 & 0.0317 & 0.0638 & 0.0525 & 0.1803 & 0.1538  \\
% No decay & 0.0394 & 0.0307 & 0.0618 & 0.0507 & 0.1773 & 0.1518  \\ 
% \bottomrule
% \vspace{-8mm}
% \end{tabular}}
% \end{table}

\subsubsection{\textbf{Effect of Update Interval.}}\label{sec:interval} 
\begin{figure}[t]
\centering
\includegraphics[width=1\linewidth]{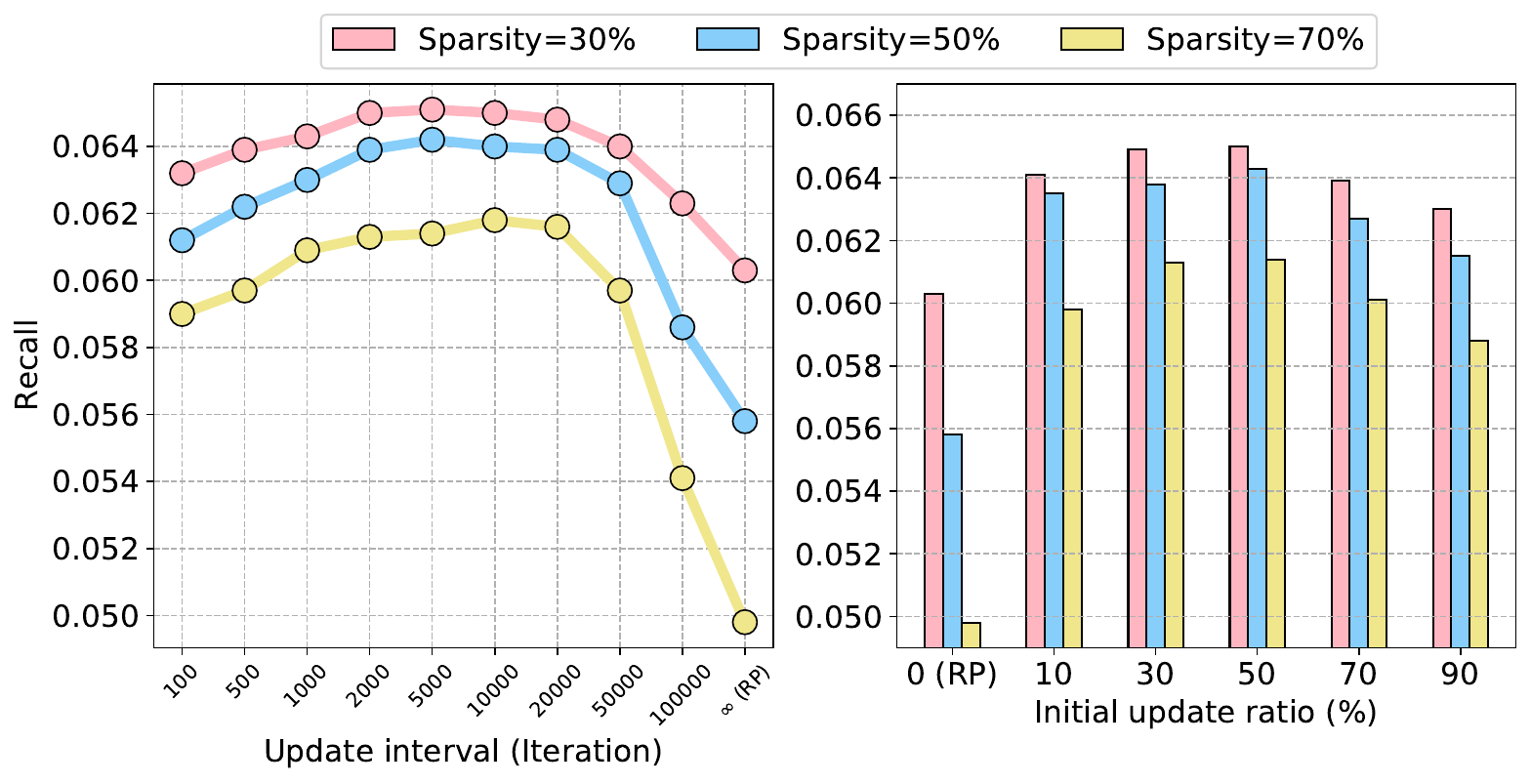}
% \vspace{-6mm}
\caption{Performance over three sparsity levels. (Left): Performance over different update intervals $\Delta T$; (Right): Performance over different initial update ratios $\rho_0$.}
\label{fig:abla}
\vspace{-4mm}
\end{figure}

To explore the effect of update intervals, we adjust ${\Delta T}$ under different sparsity levels.
${{\Delta T}=\infty}$ denotes no exploration stage during training, so DSL degenerates into RP.
From the results in Figure \ref{fig:abla} (\textit{Left}), we find that as ${\Delta T}$ increases, the performance first increases and then decreases.
On the one hand, when ${\Delta T}$ is too small (\eg $<$ 1000), the update becomes too frequent. 
Model training with the new structure does not fully converge, which leads to inaccurate judgments of parameter importance.
On the other hand, when ${\Delta T}$ is too large (\eg $>$ 50000), the model has fewer opportunities to explore more excellent structures, which leads to suboptimal performance.
Hence, we set ${\Delta T}$ to 2000$\sim$20000 in our implementations.

\begin{table}[t]
\centering
\caption{Comparisons with different decay functions.}
\vspace{-4mm}
\renewcommand\arraystretch{0.9}
\label{table:decay}
\begin{tabular}{lcccccc}
\toprule
\multirow{2}{*}{Type}   & \multicolumn{2}{c}{Amazon-book} & \multicolumn{2}{c}{Yelp2018} & \multicolumn{2}{c}{Gowalla} \\
\cmidrule(r){2-3} \cmidrule(r){4-5} \cmidrule(r){6-7}
    & Recall & NDCG & Recall & NDCG & Recall & NDCG \\
\toprule
Cosine   & 0.0406 & 0.0316 & 0.0641 & 0.0527 & 0.1812 & 0.1546  \\ 
Linear   & 0.0409 & 0.0317 & 0.0638 & 0.0525 & 0.1803 & 0.1538  \\
No decay & 0.0394 & 0.0307 & 0.0618 & 0.0507 & 0.1773 & 0.1518  \\ 
\bottomrule
\vspace{-4mm}
\end{tabular}
\end{table}

\subsubsection{\textbf{Effect of Initial Update Ratio.}} 
In DSL, $\rho_0$ refers to the initial ratio of pruning and growth in the exploration stage. 
% To explore the effects, we adjust $\rho_0$ from small to large.
In order to investigate its impact on DSL, we vary the value of $\rho_0$ across a range of small-to-large values
Please note that $\rho_0=0$ means no update during training, so DSL also degenerates into RP.
The results are depicted in Figure~\ref{fig:abla} (\textit{Right}).
We observe that as $\rho_0$ increases, the performance also shows a trend of increasing first and then decreasing.
When $\rho_0$ is too small (\eg $<$ 0.1), the scope of each exploration will be smaller, making it difficult to obtain a better model structure for each update.
If we set $\rho_0$  too large (\eg $>$ 0.7), it will also lead to poor performance.
Too large adjustments will greatly influence the stability of training, such as destroying more critical weights or adding more trivial weights.
In practice, we usually set $\rho_0$ in the range of $0.3\sim0.5$. 
Within this range, DSL can achieve an appropriate degree of dynamic adjustment, enabling the model to remove/discover an appropriate amount of trivial/critical parameters during training. 
\subsubsection{\textbf{Effect of Update Decay Function.}}\label{sec:update}
To substantiate the rationality of our choice of cosine annealing as the update decay function, we compare its performance with that of two alternative decay schedules: linear decay and no decay (i.e., where the update ratio remains constant throughout training).
From the results in Table~\ref{table:decay}, we can find that updating with no decay makes it difficult to learn effective parameters.
With each exploration, gradually reducing the update ratio is conducive to model convergence.
Compared with the linear function, using the cosine annealing function as the decay function can make the update ratio drop more smoothly and bring more stable performance.

% \begin{figure}[t] 
% \centering
% % \vspace{-0.35cm} %设置与上面正文的距离
% % \subfigtopskip=2pt %设置子图与上面正文或别的内容的距离
% % \subfigbottomskip=2pt %设置第二行子图与第一行子图的距离，即下面的头与上面的脚的距离
% % \subfigcapskip=-5pt %设置子图与子标题之间的距离
% \subfigbottomskip=0pt
% \subfigcapskip=0mm
% \subfigure[]{
% \label{fig:spr1}
% \includegraphics[width=0.40\linewidth]{Figs/popu_bar.pdf}}
% \hspace{1mm}
% \subfigure[]{
% \label{fig:spr2}
% \includegraphics[width=0.40\linewidth]{Figs/visual.pdf}}
% \vspace{-4mm}
% \caption{(a) Sparsity distribution of the embeddings. Larger GroupIDs indicate more popular users or items. (b) Embedding visualization with different popularity-levels. Darker color denotes larger weight magnitude and white color denotes the pruned weights.}
% \label{fig:cold}
% \vspace{-4mm}
% \end{figure}

\begin{figure}[t]
    \centering
    \subfloat{
        \includegraphics[width=0.48\linewidth]{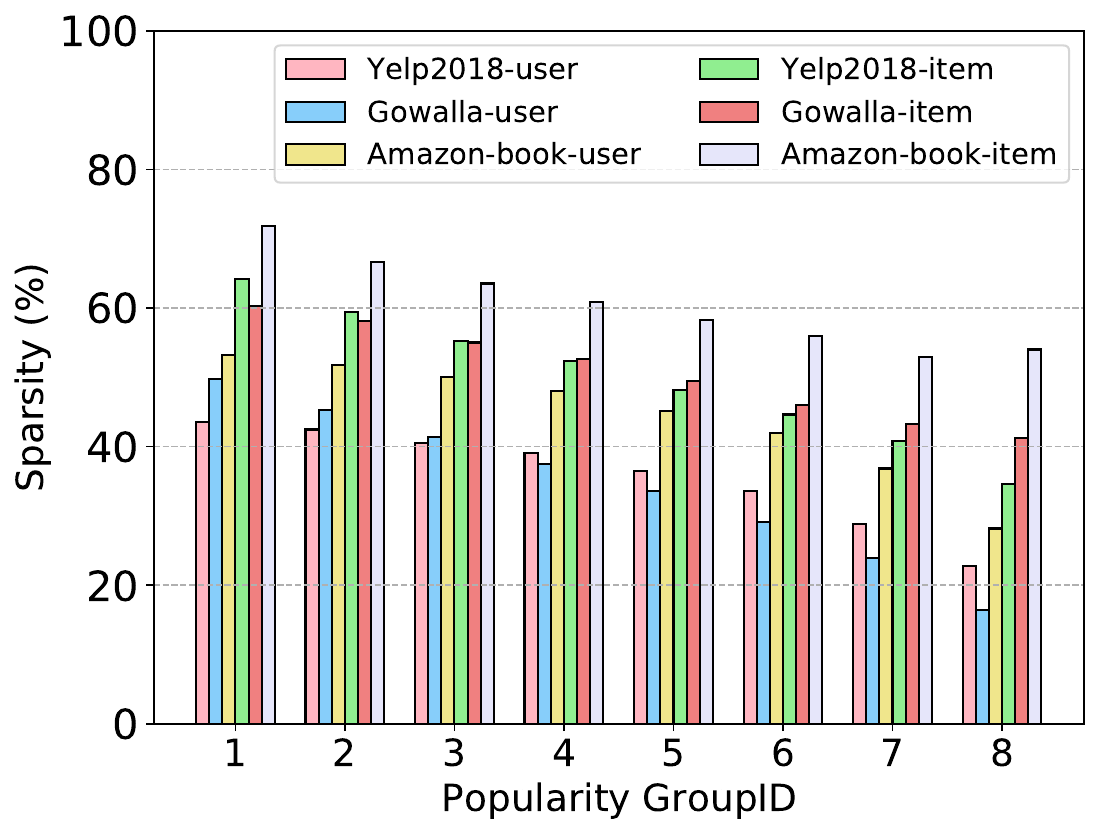}  
        }\hfill
    \subfloat{
        \includegraphics[width=0.48\linewidth]{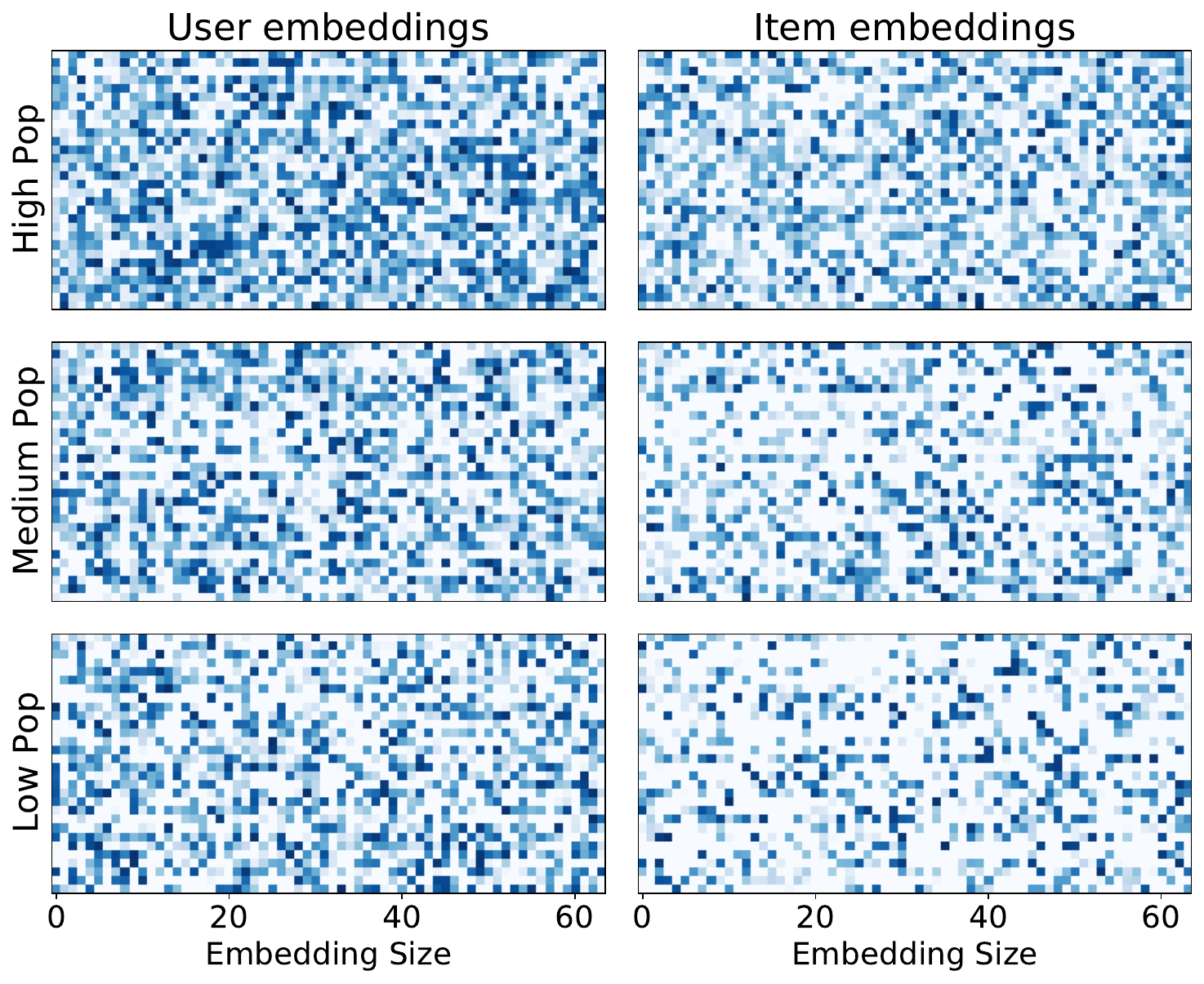}
        }
    \caption{(Left): Sparsity distribution of the embeddings. Larger GroupIDs indicate more popular users or items. (Right): Embedding visualization with different popularity-levels. Darker color denotes larger weight magnitude and white color denotes the pruned weights.}
    \label{fig:vi}
\end{figure}

% \begin{figure}[t]
% \centering
% \includegraphics[width=0.6\linewidth]{Figs/popu_bar.pdf}
% \vspace{-2mm}
% \caption{Sparsity distribution of the embeddings. Larger GroupIDs indicate more popular users or items.}
% \label{fig:pop}
% \vspace{-4mm}
% \end{figure}

\subsection{Analysis and Visualization (RQ4)}
\subsubsection{\textbf{Sparsity Distribution Analyses and Visualizations.}}
To study whether DSL can learn meaningful statistical regularities or patterns of sparsity, we conduct further analyses and visualizations about the sparse embeddings.
Firstly, we sort and uniformly group users and items by popularity, and then count the average sparsity of the embeddings within each group, as shown in Figure \ref{fig:vi} (\textit{Left}). 
To illustrate the sparsity distribution more intuitively, we visualize the sparse embeddings with different popularity levels, as shown in Figure \ref{fig:vi} (\textit{Right}).
We can observe that as the popularity increases, the sparsity of the embeddings learned by DSL gradually decreases.
DSL can indeed discover interesting patterns: active users and popular items may carry more information, thereby requiring denser embeddings,and tend to heavily prune these over-parameterized embeddings to better preserve the performance.
% We can intuitively observe that inactive users and unpopular items are more sparse.
% It indicates that these users or items may contain more redundant information. 

% \subsubsection{\textbf{Sparsity Distribution Analyses.}}
% To study whether DSL can learn meaningful statistical regularities or patterns of sparsity, we conduct further analyses about the sparse embeddings.
% Firstly, we sort and uniformly group users and items by popularity, and then count the average sparsity of the embeddings within each group.
% As shown in Figure \ref{fig:pop}, we can observe that as the popularity increases, the sparsity of the embeddings learned by DSL gradually decreases.
% This phenomenon illustrates that DSL can indeed discover some interesting patterns: active users and popular items may carry more information, thereby requiring denser embeddings.

\subsubsection{\textbf{Convergence Analyses}}
\begin{figure}[t]
\centering
\includegraphics[width=1\linewidth]{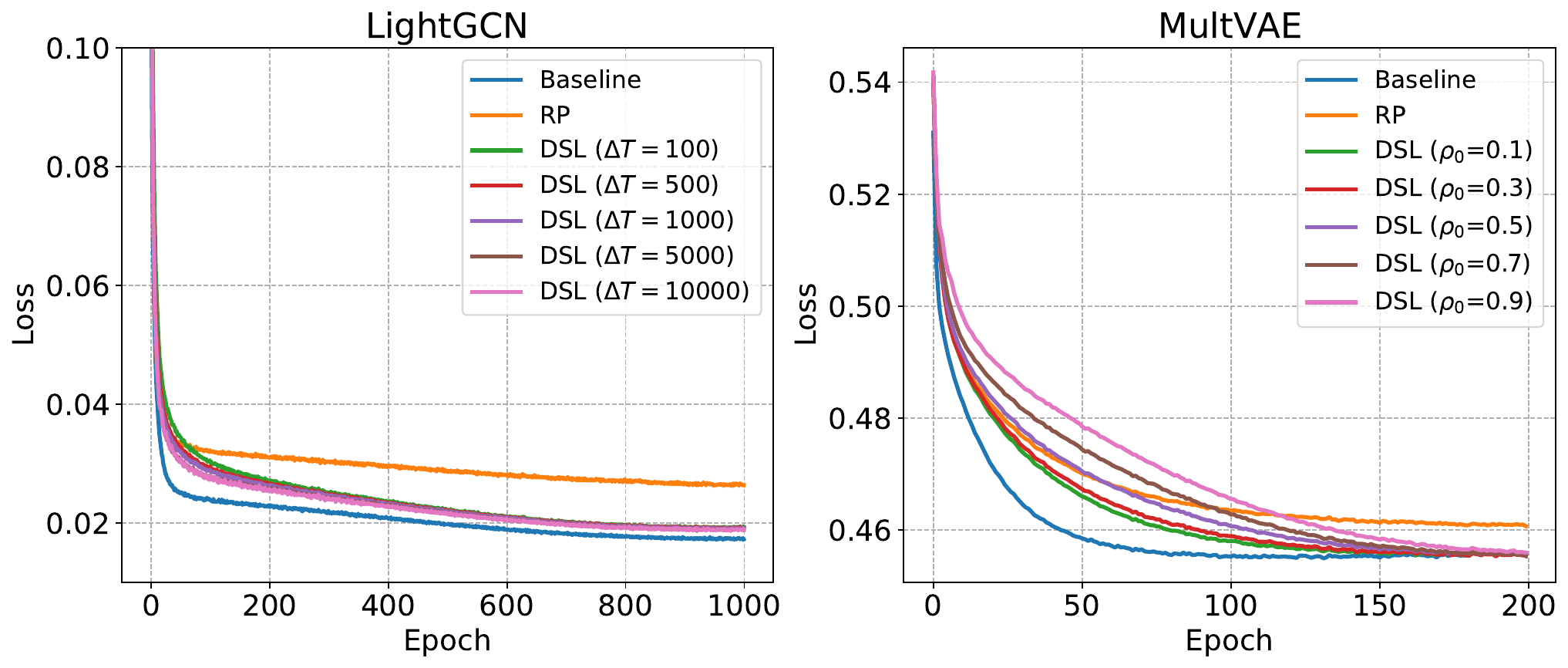}
\vspace{-4mm}
% \caption{Convergence comparisons.}
\caption{Convergence comparisons of training loss for the baseline model, RP, and DSL on different initial update ratios $\rho_0$ and update intervals $\Delta T$.   }
\label{fig:conver}
\vspace{-4mm}
\end{figure}

To better illustrate the influence of DSL on training convergence, we modify its parameters - initial update ratios $\rho_0$ and update intervals $\Delta T$ - and plot the corresponding loss decline curves for LightGCN and MultVAE models in Figure~\ref{fig:conver}.
At the beginning of training, DSL shows a rapid drop in training loss at a similar speed to the baseline and RP. 
However, as the number of iterations increases, the curves generated by DSL with varying update ratios and intervals exhibit slower convergence rates at higher update ratios or smaller update intervals. 
Compared to RP, DSL enables periodic parameter updates that facilitate faster convergence to lower loss values without impeding model convergence.
Furthermore, compared to the baseline, while DSL does exhibit a slightly slower convergence speed compared to the baseline, it achieves comparable loss levels under nearly identical training epochs.

\section{Related Work}\label{sec:rw}
   
\noindent\textbf{Model Compression} \cite{wang2019eigendamage,yan2021learning,zhou2023opportunities,molchanov2019importance,evci2020rigging,qu2022single} aims to obtain lightweight models by removing redundant weights.
It becomes an effective solution to reduce model size in the fields of computer vision~\cite{chen2020gans,sui2022towards}, graph learning~\cite{JCST-2206-12583,sui2022causal,sui2022adversarial,sui2023unleashing,fangjf1,fangjf2,chen2021unified}, and natural language processing~\cite{chen2020mmea,chen2022entity,chen2022msnea}.
Recently, the lottery ticket hypothesis \cite{frankle2018lottery,wang2022exploring} becomes an effective strategy to guide the model pruning.
It states that dense randomly initialized networks contain sparse subnetworks (\aka winning tickets) that can be trained in isolation to achieve comparable performance.
LTH-MRS \cite{wang2022exploring} adopts an iterative pruning strategy to find the winning tickets in recommendation models, while still suffering from the expensiveness of multi-round retraining.
PEP \cite{liu2021learnable} proposes learnable pruning thresholds in model training, while the unstable learning process of thresholds makes it difficult to keep a controllable training budget and a stable performance.
Distinct from them, we adjust the sparsity distribution of model weights while sticking to a fixed parameters budget, which can achieve both training and inference efficiency.

\vspace{5pt}\noindent\textbf{Efficient Recommendation} is drawing widespread attention, due to the emergence of large-scale recommendation models \cite{wang2022exploring, chen2021towards}. 
Knowledge Distillation (KD) \cite{kweon2021bidirectional,kang2021topology,kang2022personalized,tang2018ranking} improves the performance of small-scale student models by distilling the knowledge from large-scale teacher models.
BKD \cite{kweon2021bidirectional} proposes a bidirectional distillation framework, which enhances the collaborations between teacher and student models. 
TKD \cite{kang2021topology} develops a topology distillation strategy, which transfers topological knowledge and guides students with diverse capacities. 
However, they need to pre-train large-scale teacher models, which still leads to expensive memory and training costs.
Furthermore, there also exists a large gap in performance between student and teacher models \cite{tang2018ranking}.
Automated Machine Learning (AutoML) \cite{automlsurvey,zheng2022automl,liu2021learnable,zhao2021autodim,zhaok2021autoemb,liu2020automated} improves inference efficiency by searching powerful and lightweight model architectures.
AutoEmb \cite{zhaok2021autoemb} adopts a DARTS-based \citep{liu2018darts} optimization strategy to decide optimal embedding dimensions for users and items.
ESAPN \cite{liu2020automated} leverages an automated reinforcement learning agent to search the embedding size of models.
However, they require to predefine a search space with human prior knowledge.
Meanwhile, the complex processes of optimization and performance estimation also result in an unaffordable computational cost.
In contrast, our strategy can effectively trim down both training and inference overheads, by simply training a lightweight sparse model in an end-to-end way.

\section{Conclusion and Future Work}
In this work, we proposed DSL, which aims to trim down both the training and inference costs for recommendation models.
Specifically, we introduced the concept of sparsity to the models, periodically and dynamically adjusted the sparsity distribution of model weights, and sticked to a fixed parameter budget throughout the entire learning lifecycle.
Different from existing solutions, DSL achieved the intriguing prospect of “end-to-end” efficiency from training to inference.
We conducted extensive experiments on diverse recommendation models with six benchmark datasets.
The experimental results demonstrated that DSL can largely reduce the training cost, inference cost, and memory, with comparable recommendation performance.
Finally, we also provided analyses and visualizations to demonstrate the effectiveness and rationality.

\section{Acknowledgments}
This research was supported in part by the National Natural Science Foundation of China (Grant No. 92370204) and Guangdong OPPO Mobile Telecommunications Co., Ltd .

\bibliographystyle{ACM-Reference-Format}
\bibliography{DSL}

\end{document}